\documentclass[pre,amsmath,amssymb,twocolumn,superscriptaddress]{revtex4}
\usepackage{amsmath,amssymb}
\usepackage{graphicx}
\usepackage{dcolumn}

\begin{document}

\title{Antipersistent energy current correlations in strong long-ranged Fermi-Pasta-Ulam-Tsingou type models}

\author{Daxing Xiong}
\email{xmuxdx@163.com}
\affiliation{MinJiang Collaborative Center for Theoretical Physics, College of Physics and Electronic Information Engineering, Minjiang University, Fuzhou 350108, China}

\author{Jianjin Wang}
\email{phywjj@foxmail.com}

\affiliation{Department of Physics, Jiangxi Science and Technology Normal University, Nanchang 330013, Jiangxi, China}
\affiliation{MinJiang Collaborative Center for Theoretical Physics, College of Physics and Electronic Information Engineering, Minjiang University, Fuzhou 350108, China}

\begin{abstract}
We study heat transfer in one-dimensional Fermi-Pasta-Ulam-Tsingou type systems with long-range (LR) interactions. The strength of the LR interaction between two lattice sites decays as a power $\sigma$ of the inverse of their distance. We focus on the strong LR regime ($0\leq \sigma \leq1$) and show that the thermal transport behaviors are remarkably nuanced. Specifically, we observe that the antipersistent (negative) energy current correlation in this regime is intricately dependent on $\sigma$, displaying a nonmonotonic variation. Notably, a significant qualitative change occurs at $\sigma_c=0.5$, where with respect to other $\sigma$ values, the correlation shows a minimum negative value. Furthermore, our findings also demonstrate that within the long-time range considered, these antipersistent correlations will eventually vanish for certain $\sigma >0.5$. The underlying mechanisms behind these intriguing phenomena are related to the crossover of two diverse space-time scaling properties of equilibrium heat correlations and the various scattering processes of phonons and discrete breathers.
\end{abstract}

\maketitle

\section{Introduction}
Recent studies have increasingly focused on Long-range (LR) interactions, prevalent from the broad scope of cosmology~\cite{LR-1} to the intricacies of nanoscience~\cite{LR-2}. These systems are characterized by interaction potentials $V(r)$ that decay with the power law
\begin{equation}
V(r) \propto \frac{1}{r^{\sigma}}
\end{equation}
with $r$ the distance between two interacting particles. These interactions lead to systems with extremely rich dynamics and thermal properties, distinctly different from systems with short-range interactions. Such systems frequently exhibit a range of unique phenomena: nonergodic behavior, weak chaos, ensemble inequivalence, long-lived non-Gaussian quasi-stationary states, phase transitions in one-dimensional (1D) contexts, nonconcave entropy, and even negative specific heat (see the review articles~\cite{LR-3,LR-4,LR-5,LR-6}). These exceptional properties render the classical Boltzmann-Gibbs statistical mechanics framework inadequate. The intriguing and complex nature of LR interacting systems makes them a compelling yet challenging field of study.

However, despite the expectation of distinct roles for the range of interactions in heat transport, most studies have focused on nearest-neighbor (NN) couplings~\cite{NNHeatreview-1,NNHeatreview-2,NNHeatreview-3}. Only recently has thermal transport in LR interacting systems garnered attention~\cite{LRHeatreview-1,LRHeatreview-2,LRHeatreview-3}. In this context, several paradigmatic models, such as the variants of LR rotor, Fermi-Pasta-Ulam-Tsingou (FPUT), lattice $\phi^4$ and harmonic models have been proposed and different ranges of LR interactions characterized by $\sigma$ have been considered~\cite{LRHeat-2016,LRHeat-2017-1,LRHeat-2017-2,LRHeat-2018,
LRHeat-2019,LRHeat-ours-1,LRHeat-2020,LRHeat-2021,LRHeat-ours-2,LRHeat-2022-1,LRHeat-2022-2,
LRHeat-2022-3,LRHeat-2023}. For both nonlinear rotor and FPUT-type models with conserved momentum, a detailed $\sigma$-dependent thermal transport has been examined~\cite{LRHeat-2016,LRHeat-2017-1,LRHeat-2017-2,LRHeat-2018,
LRHeat-2019,LRHeat-ours-1,LRHeat-ours-2}. The LR-rotor model was found to exhibit two distinct phases, i.e., an insulator phase in the strong LR regime (where $0< \sigma< 1$) and a conducting phase when $\sigma > 1$~\cite{LRHeat-2016}. In particular, in the strong LR regime, an unusual flat temperature profile, similar to the behaviors displayed in integrable systems was observed~\cite{LRHeat-2016}. This phenomenon is seemingly to be linked to integrability. However, further investigation~\cite{LRHeat-2018} revealed that it is influenced by a parallel energy transport mechanism, but not integrability. Additionally, the insulator phase's subdiffusive energy transport mechanism in the mean-field case (with $\sigma = 0$) was confirmed numerically in Ref.~\cite{LRHeat-2017-2}.

In studies of the LR-FPUT model, LR interactions can be added in two ways, i.e., LR quartic potential terms~\cite{LRHeat-2017-1,LRHeat-ours-1,LRHeat-2021,LRHeat-ours-2} and a combination of both LR quadratic and quartic terms~\cite{LRHeat-2018,LRHeat-2019}. These studies have revealed more interesting transport properties compared to the LR-rotor model. A notable discovery~\cite{LRHeat-2017-1} is the behavior at $\sigma=2$, where thermal conductivity $\kappa$ shows an almost linear divergence with system size $L$ ($\kappa \sim L^\alpha$ with $\alpha=1$) and the temperature profile has a negligible slope. This ballistic-like transport hints at some yet unknown integrable limit of the model at $\sigma=2$~\cite{LRHeatreview-2,LRHeatreview-3}. However, this LR interacting system is a nonintegrable one. Our subsequent study~\cite{LRHeat-ours-1} carefully considered the reservoir and boundary effects and observed a superdiffusive transport, instead, with a significantly high divergent exponent $\alpha \approx 0.7$, much larger than the $\alpha \approx 0.3-0.5$ typical of short-range interacting systems with nearest-neighbor (NN) couplings only. This peculiar superdiffusive transport at $\sigma=2$ was further characterized by a slower decay in energy current correlation and linked to a weaker nonintegrability mechanism and the presence of a special type of traveling discrete breathers (DBs) in the system that enhance thermal transport~\cite{LRHeat-ours-1}. Although this new type of superdiffusive energy transport was later doubted by Ref.~\cite{LRHeat-2021}, the collective body of works~\cite{LRHeat-2017-1,LRHeat-2018,LRHeat-ours-1,LRHeat-2021} do confirm the unique properties observed at $\sigma=2$.

Beyond $\sigma=2$, it was initially posited that the LR-FPUT model exhibits superdiffusive transport across all other $\sigma$ values~\cite{LRHeat-2017-1}. However, the parallel energy transport mechanism, proposed in~\cite{LRHeat-2018} is likely taking its effects within $0<\sigma<1$ and a detailed scaling analysis of equilibrium correlation for $1<\sigma<3$ indicates a unique non-standard, non-local effective hydrodynamics, which differ from those observed in short-range interacting systems with NN couplings only~\cite{LRHeat-2019}. Therefore, it is quite strange that for $0<\sigma<1$, the LR-FPUT model demonstrates superdiffusive transport, while the LR-rotor model shows an insulating phase, indicative of subdiffusive energy transport. This disparity raises questions about the striking differences in transport properties between these two models~\cite{LRHeat-2017-2}. To clarify this disparity, our recent study~\cite{LRHeat-ours-2} has unequivocally corroborated the existence of subdiffusive energy transport in the mean-field limit of the LR-FPUT model, akin to that observed in the LR-rotor model.

In view of these developments, the $2020$ study by Tamaki and Saito~\cite{LRHeat-2020} aimed to derive a clear-cut analytical result that would complement the existing numerical findings. They conducted an analytical investigation of thermal transport properties in a linear LR-harmonic system, which involves stochastic momentum-exchange dynamics with both conserved and nonconserved momenta. Their idea is to calculate the equilibrium energy current correlation function, and then applying the Green-Kubo formula to determine thermal conductivity. However, their calculations were restricted to the weak LR regime, specifically for $\sigma \geq 2$ in momentum-conserving systems, and $\sigma \geq 3/2$ in momentum-nonconserving systems. For sufficiently strong LR systems, the divergence of the current correlation hampers their calculations. Despite this challenge, the work of Tamaki and Saito~\cite{LRHeat-2020} has spurred further recent studies on systems without involving stochastic dynamics, such as the momentum-nonconserving LR-$\phi^4$ model~\cite{LRHeat-2022-1}; the momentum-conserving LR-harmonic systems in the mean-field limit~\cite{LRHeat-2022-2, LRHeat-2022-3} and investigations into the strong LR regime beyond the mean-field case both using the nonequilibrium Green's function formalism~\cite{LRHeat-2023}.

In this work we study numerically the thermal transport properties of LR-FPUT model, specifically focusing on the strong LR regime beyond the mean-field case ($0 \leq \sigma \leq 1$)---a domain not yet addressed by current analytical methods. Instead of examining the system size-dependent thermal conductivity as in previous studies~\cite{LRHeat-2016,LRHeat-2017-1,LRHeat-2018}, our study concentrates on calculating the equilibrium energy current correlation function. Since in Ref.~\cite{LRHeat-ours-2} we have observed the subdiffusive energy transport and its antipersistent energy current correlation in the LR-FPUT model in the mean-field case, and our results are distinct from the superdiffusive transport observed in~\cite{LRHeat-2017-1}, in the present work we also wish to explore further how this antipersistent energy current correlation would change with the variation of $\sigma$.

The remainder of our work is organized as follows: Section II delineates the specific LR-FPUT model under study within the strong LR regime, together with the details of the numerical methods employed. In Section III, we present our findings on the $\sigma$-dependent energy current correlation in this regime. Section IV is dedicated to unraveling the underlying mechanisms of the observations, with a focus on analyzing the scaling behavior of the equilibrium heat correlations. Section V extends our investigation to the microscopic scattering dynamics of phonons and DBs, aiming to provide more insight. Finally, we conclude our work in Sec. VI with a short summary.

\section{Model and method}
We consider a 1D FPUT-type LR interacting system of $N$ particles with Born-von Karman periodic boundary condition~\cite{BornvonKarman} whose dynamics is governed by the Hamiltonian:
\begin{equation}
H=\sum_{i=1}^N \left[\frac{p_{i}^{2}}{2}+ \frac{1}{2} (x_{i+1}-x_i)^2 + \frac{1}{4} \sum_{r=1}^{\frac{N}{2}-1} \frac{(x_{i+r}-x_i)^4}{r^{\sigma}}  \right].
\label{HH}
\end{equation}
In this Hamiltonian, $x_{i}$ and $p_{i}$ are two canonically conjugated variables with $i$ the index of the particle; all other relevant quantities like the particle's mass and the lattice constant are dimensionless and set to be unity.
The periodic boundary conditions used make our system like a ring which ensures that our study does not suffer from any uncertain effects caused by the open boundary conditions. The interparticle interaction has two separate terms as our previous works~\cite{LRHeat-ours-1,LRHeat-ours-2}. The NN couplings are added to the quadratic potential term and the LR interactions are included in the quartic potential term. In the LR potential term, $r^{-\sigma}$ represents the interaction strength of the $i$th particle with its $r$th neighborings with $\sigma$ the range value of LR exponent. Such a setup of the LR-FPUT model including the LR quartic potential term only has been found to showing more interesting dynamic phenomena than that by a combination of both the LR quadratic and quartic potential terms~\cite{JianWang2022,JianWang2022cite}, and therefore, in the present work we will focus on this case only.

We do not include the Kac scaling factor $\tilde{N}$ into the LR potential term. This is because this Kac factor was designed to restore system's extensivity as increasing system size, but it does not help improve system's nonadditivity~\cite{LR-4}. It only constructs an ``artificial'' extensive system, but the cost for thermal transport is that both the phonon's group velocity and the strength of nonlinearity should depend on $N$, which is an unwanted effect for thermal transport studies~\cite{LRHeat-ours-1}. In addition, the dynamical time that relaxed to equilibrium should be $\sqrt{\tilde{N}}$ longer if we include $\tilde{N}$~\cite{Tamarit2000}. In fact, several subsequent studies~\cite{LRHeat-2021,LRHeat-2022-2} have confirmed that there are not obvious difference of transport between systems of including or not including $\tilde{N}$. We would also like to note that a system like Hamiltonian~(\ref{HH}) at $\sigma=2$ but not including $\tilde{N}$ has been shown to have a special symmetric feature and support a type of free-tail traveling DBs~\cite{Doi2016,LRHeat-ours-1}. This is also why we choose not to include the Kac scaling factor.

Finally, as mentioned in Introduction, our study will concentrate on the transport in the strong LR regime beyond the mean-field case ($0 \leq \sigma \leq 1$). Toward this end, we shall study the system's equilibrium energy current time ($t$) autocorrelation, defined by
\begin{equation}
C_{JJ}(t)=\langle J_{\rm tot} (t) J_{\rm tot}(0) \rangle,
\end{equation}
with
\begin{equation}
J_{\rm tot}=\sum_{i=1}^{N} p_i \left[(x_{i+1}-x_i)+ \sum_{r=1}^{\frac{N}{2}-1} \frac{(x_{i+r}-x_i)^3} {r^{\sigma}}\right]
\label{JJ}
\end{equation}
the total heat current along the system. Note that this energy current of Eq.~\ref{JJ} is a LR form which is different from the definition in~\cite{LRHeat-2017-1} by using the harmonic lead sites. The calculation of our definition is then more challenging.

For diffusive and superdiffusive transports, $C_{JJ}(t)$ is related to the thermal conductivity $\kappa$ through the Green-Kubo formula~\cite{LRHeat-2020}
\begin{equation}
\kappa=\lim_{\tau \rightarrow \infty} \lim_{N \rightarrow \infty} \frac{1}{k_B N T^2} \int_{0}^{\tau} C_{JJ}(t) dt,
\end{equation}
where $k_B$ is the Boltzmann constant and $T$ is the systems's equilibrium temperature. However, in systems within the strong LR regime that may support subdiffusive transport, the energy current correlation $C_{JJ}(t)$ is ill defined~\cite{LRHeat-2020}. It shows an antipersistent correlation with negative values of correlations~\cite{LRHeat-ours-2}.

To numerically calculate $C_{JJ}(t)$, for each $\sigma$, we fix the total system size $N=4096$ and prepare a fully thermalized system under the equilibrium averaged temperature $T=0.5$. The system is evolved by velocity-Verlet algorithm~\cite{Verlet} with a small time step $ 0.01$, that guarantees energy conservation with a relative accuracy of $O(10^{-5})$. We adopt a Fast Fourier Transform~\cite{FFT} algorithm to accelerate our computations. This helps avoid $O(N^2)$ operations in calculating forces at each time step in our system. We utilize an ensemble of size about $8 \times 10^9$ to derive the correlation.

\section{Range-dependent energy current correlations}
\begin{figure}[!t]
\vskip-0.2cm
\includegraphics[width=8.8cm]{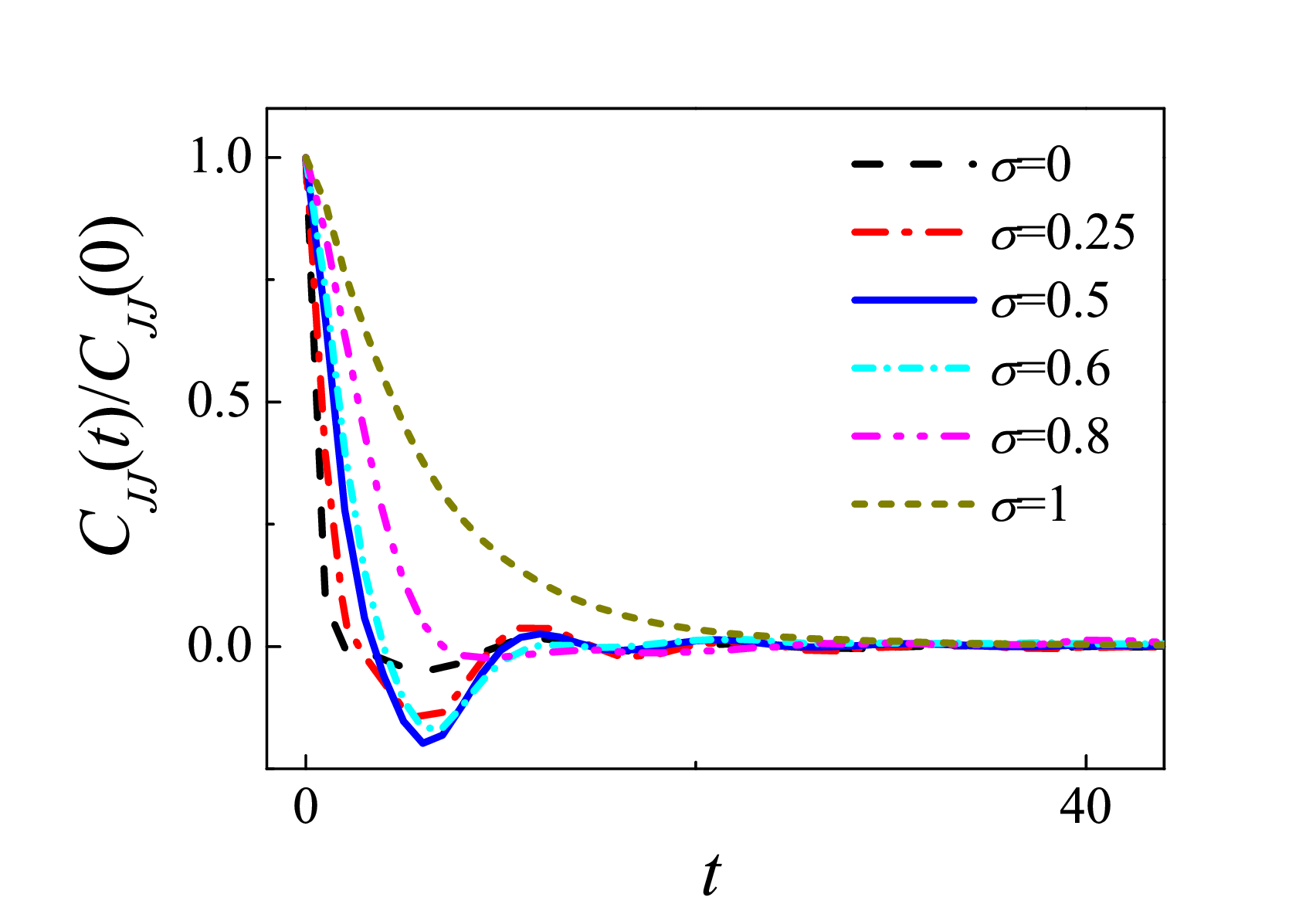}
\vskip-0.3cm
\caption{The equilibrium heat current autocorrelation $C_{JJ} (t)$ versus $t$ for several $\sigma$ within the LR strong regime (with $0 \leq \sigma \leq 1$). Here for a better visualization, the correlation values have been rescaled between $0$ and $1$ by dividing them with $C_{JJ} (0)$.} \label{fig:1}
\end{figure}
Figure~\ref{fig:1} depicts $C_{JJ}(t)$ as a function of the time lag $t$ for
various LR exponent $\sigma$ within $0 \leq \sigma \leq 1$. We see that for $\sigma=0$, i.e., the mean-field case, $C_{JJ}(t)$ shows a negative minimum at a short time and finally turns to zero in a relative long time, just like what was observed in Ref.~\cite{LRHeat-ours-2}, i.e., the interesting antipersistent correlation. As $\sigma$ increases, a intriguing \emph{nonmonotonic} change takes place, i.e., for $0 < \sigma \leq 0.8$, even though the antipersistent correlations are always present,
the negative minimum now first becomes smaller, and reaches its smallest value at $\sigma_c=0.5$, and then increases. After about $\sigma \approx 0.8$, a final exponential-like decay of $C_{JJ}(t)$ emerges, indicating the absence of the antipersistent correlation. We stress that this nonmonotonic variation of $C_{JJ}(t)$ that from presence to absence of antipersistent correlations is indeed a new intriguing phenomenon, which is quite distinct from the previously observed range-dependent correlations in the mean-field case (see Fig.~5 in Ref.~\cite{LRHeat-ours-2}). We also note that such a variation of current/velocity antipersistent correlations has never been explored in the particle subdiffusion models~\cite{PhysRP2015,Sub-4,Sub-5}, thus it may shed new light on further studies of particle subdiffusion. Finally, these antipersistent energy current correlations are an obvious sign of subdiffusive energy transport, strongly suggesting that the previous observation of superdiffusive thermal transport in the LR-FPUT model in Ref.~\cite{LRHeat-2017-1} is not the real transport behavior of the system. Therefore, both the LR-FPUT and LR-rotor models within the strong LR regime beyond the mean-field case should follow subdiffusive energy transport.

\section{Space-time scaling properties of equilibrium heat correlations}
\begin{figure}[!t]
\vskip-0.2cm
\includegraphics[width=8.8cm]{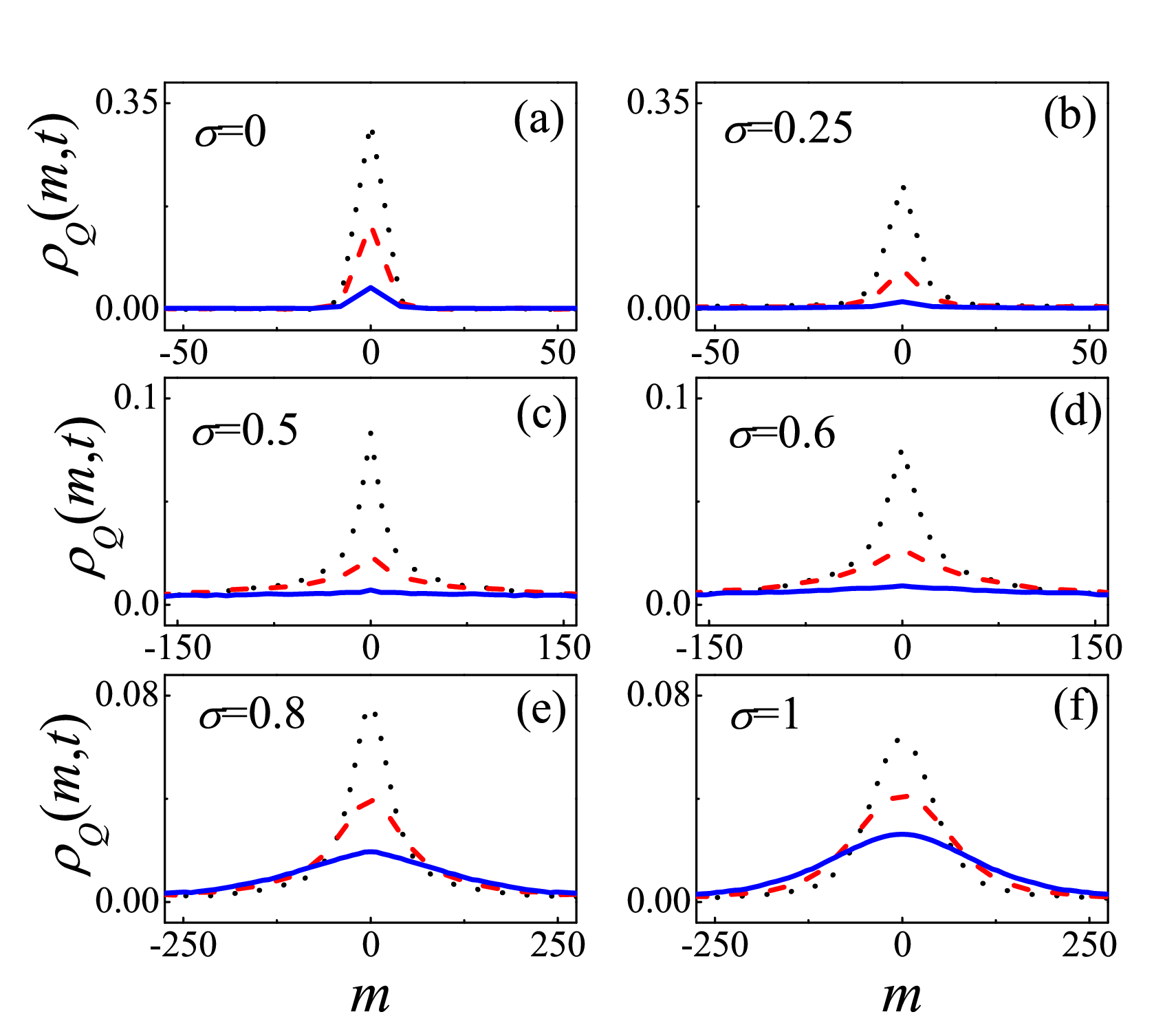}
\vskip-0.3cm
\caption{The equilibrium spatiotemproal correlation function of heat $\rho_Q(m,t)$ for several $0\leq \sigma \leq 1$ for three typical long times: $t = 1000$ (dotted), $t = 2000$ (dashed), and $t = 4000$ (solid).} \label{fig:2}
\end{figure}
To understand such nonmonotonic variation of energy current correlations, we next study the space-time scaling properties of the equilibrium spatiotemproal correlation function of the local thermal energy, defined by~\cite{LRHeat-ours-1,Zhao2013}
\begin{equation}
\rho_{Q}(m,t)=\frac{\langle \Delta Q_{l+m}(t) \Delta Q_{l}(0) \rangle}{\langle \Delta Q_{l}(0) \Delta Q_{l}(0) \rangle},
\label{heat}
\end{equation}
where $\langle \cdot \rangle$ represents the spatiotemporal average; $l$ labels a coarse-grained bin's number (in practice we set each bin of $8$ particles); $Q_l(t)=E_l(t)-\frac{(\langle E \rangle + \langle F\rangle) g_l(t)}{\langle g \rangle}$ is the local thermal energy with $g_l(t)$ the particle number density, $E_l(t)=\sum_{i} E_i(t)$ the energy density within the $l$th bin [where $E_i=\frac{p_{i}^{2}}{2}+ \frac{1}{2} (x_{i+1}-x_i)^2+ \frac{1}{4} \sum_{r=1}^{\frac{N}{2}-1} \frac{(x_{i+r}-i_k)^4}{r^\sigma}$], and $F_l(t)(\langle F\rangle \equiv 0)$ the pressure density, respectively.

We note that due to the translational invariance, $\rho_{Q}(m,t)$ only depends on the relative distance $m$. Therefore, if known $g_l(t)$ and $E_l(t)$ in the $l$th bin, and $F_l(t)$ exerted on the $l$th bin at each time $t$, the detailed calculation of $\rho_Q(m,t)$ is quite similar to that of $C_{JJ}(t)$. Here, we consider the same system size, set the same averaged temperature, and utilize the same ensemble of size.

Figure~\ref{fig:2} depicts $\rho_Q (m,t)$ for three long times for several $0\leq \sigma \leq 1$. First, for all of $\sigma$ considered, it shows
a relatively slow spread of heat [all $\rho_Q (m,t)$ for a time $t=4000$ only cover few lattice sites, which are much smaller than the system size], indicating the possible mechanism of subdiffusive thermal transport as shown in Fig.~\ref{fig:1}. Again, this is distinct from the superdiffusive transport observed in Ref.~\cite{LRHeat-2017-1}. Another information is that with the increase of $\sigma$, especially for $\sigma=0.8$ and $\sigma=1$ [see Figs.~\ref{fig:2}(e,~f)], $\rho_Q (m,t)$ appears to be delocalized, indicating the absence of subdiffusive energy transport. This is consistent with the observation of energy current correlations in Fig.~\ref{fig:1}.
\begin{figure}[!t]
\vskip-0.2cm
\includegraphics[width=8.8cm]{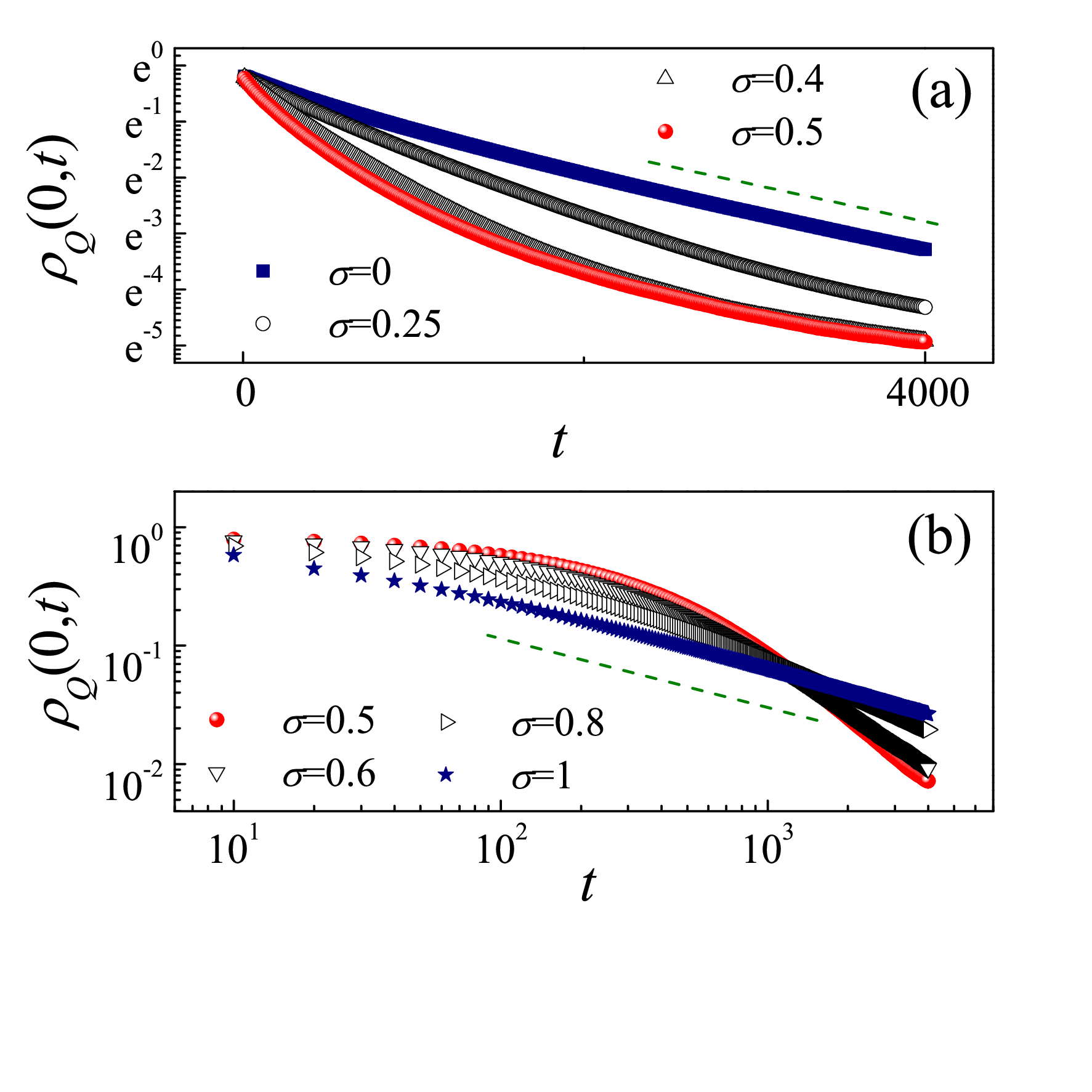}
\vskip-0.3cm
\caption{(a) A single log plot of $\rho_Q (0,t)$ versus $t$ for describing an exponential decay for $\sigma=0$, $0.25$, $0.4$ and $0.5$; (b) A log-log plot of $\rho_Q (0,t)$ versus $t$ for describing a power-law decay for $\sigma=0.5$, $0.6$, $0.8$ and $1$. The two dashed lines are a guider for the two distinct laws.} \label{fig:3}
\end{figure}

We now try to understand why $\sigma_c=0.5$ is a special point. We will not study the mean squared displacement of $\rho_Q(m,t)$ versus $t$ here because of the relatively large fluctuations of $\rho_Q(m,t)$ for a large distance $m$. Instead, we will focus on the decay of the central peak $\rho_Q(0,t)$ of $\rho_Q(m,t)$. Such a decrease of $\rho_Q(0,t)$ versus $t$ has been used to infer the long-time asymptotic behavior of heat spread for the short-range systems with NN couplings only~\cite{DX2018,DX2017-1,DX2017-2} and for the LR-FPUT models~\cite{LRHeat-2017-1,LRHeat-ours-1,LRHeat-2021} in the regimes of ballistic, superdiffusive, and diffusive transport. In those cases, if $\rho_Q(0,t) \sim t^{-1/\gamma}$, based on the scaling formula of the single particle L\'{e}vy walk model~\cite{Report2015} we have
\begin{equation} \label{Scaling}
t^{1/\gamma} \rho_Q(m,t) \simeq \rho_Q(t^{-1/\gamma}m,t).
\end{equation}
Then, $\gamma=1$, $1<\gamma<2$, and $\gamma=2$ correspond to ballistic, superdiffusive, and diffusive transport, respectively, where one finds $1/2 \leq 1/ \gamma \leq 1$. However, for subdiffusive thermal transport considered here, one does not have such a scaling law of $\rho_Q(0,t) \sim t^{-1/\gamma}$. Fortunately, in Ref~\cite{LRHeat-ours-2} we for the first time revealed that in the LR-FPUT model of the mean-field case ($\sigma=0$) following subdiffusive thermal transport, a \emph{two-stage} exponential decay law $\rho_Q(0,t) \sim \exp (-\frac{t}{\tau})$ is available. Therefore, bearing the results of Figs.~\ref{fig:1} and~\ref{fig:2} in mind, one may conjecture a crossover between these two distinct laws for $\rho_Q(0,t)$ versus $t$ as $\sigma$ increases.

To provide evidence for our conjecture, Fig.~\ref{fig:3} presents the results of $\rho_Q(0,t)$ versus $t$ in two distinct ways: (a) a single-log plot and (b) a log-log plot. As can be seen, both results of $\sigma=0$ and $\sigma=1$ for long times can be well fitted by a line. This indicates that there is indeed a crossover between the two distinct laws, which characterizes different transport regimes. Furthermore, since $\sigma_c=0.5$ lies between the extremes of $\sigma=0$ and $\sigma=1$, this finding provides preliminary evidence supporting the speciality of $\sigma_c=0.5$, as indicated by the antipersistent energy current correlations depicted in Fig.~\ref{fig:1}.
\begin{figure}[!t]
\vskip-0.2cm
\includegraphics[width=8.8cm]{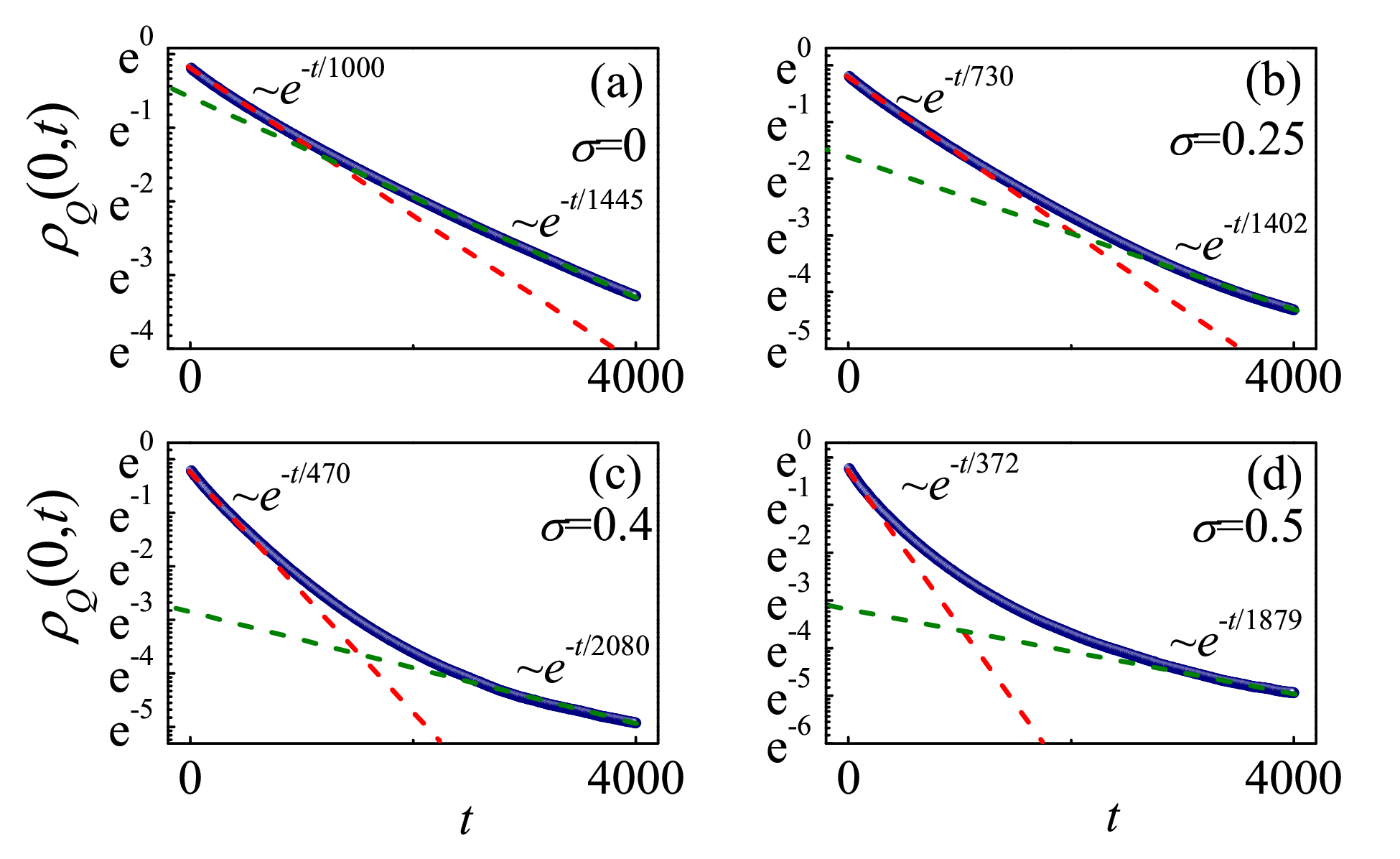}
\vskip-0.3cm
\caption{$\rho_Q(0,t)$ versus $t$ in a single-log plot showing the detailed two-stage exponential decay for (a) $\sigma=0$, (b) $\sigma=0.25$, (c) $\sigma=0.4$, and (d) $\sigma=0.5$.} \label{fig:4}
\end{figure}

From Fig.~\ref{fig:1} the antipersistent energy current correlations mainly occur in a relatively short time, it is thus necessary to examine more detailed variation of the scaling of $\rho_Q(0,t)$ in a short time. Based on the current data with a time up to $t=4000$, we have tried both plots of single-log and log-log. We find that the single-log plot is helpful to highlight the short-time behavior. Therefore, in the following we will first present the details of the two-stage exponential decay behavior of $\rho_Q(0,t)$ for $\sigma \in [0,0.5]$, then focus on the results on the vicinity of $\sigma_c=0.5$, and finally provide the power-law decay of $\rho_Q(0,t)$ for $\sigma \in (0.5,1]$. Figure~\ref{fig:4} depicts the two-stage exponential decay behavior for four typical $\sigma$ values within $[0,0.5]$. As to this figure, we first recall the known results of $\sigma=0$. Indeed, similar to the observations for the equilibrium spatiotemporal correlations of total energy in~\cite{LRHeat-ours-2}, we see a perfect two-stage exponential decay characterized by two distinct timescales: $\tau_1 \simeq 1000$ for short times and $\tau_2 \simeq 1445$ for long times [see~Fig.~\ref{fig:4}(a)]. In particular, this perfect two-stage exponential decay has been conjectured to related to the two different energy relaxation induced by the system's surviving DBs, before and after phonon-DBs scattering, respectively. In this conjecture, due to the significantly stronger strength of DBs relative to phonons in the mean-field scenario, phonon-DB scattering acts as a transient process which leads to a rapid switching between the two exponential decays as mentioned in~\cite{LRHeat-ours-2} and also shown in Fig.~\ref{fig:4}(a). This may be the reason why we observe a relatively small value of $|C_{JJ}(t)|$ at its negative minimum for $\sigma=0$ if compared to other $\sigma \in [0,0.5]$ [see Fig.\ref{fig:1}].

If we acknowledge the reasonability of the above conjecture, let us next discuss the results depicted in Figs.~\ref{fig:4}(b)-(d). The key distinction if compared to $\sigma=0$ is that the switching between the two exponential decays now becomes progressively slower. This may imply a prolonged persistence of phonon-DBs scattering, resulting in an increased reflection of heat flux and consequently leading to a larger value of $|C_{JJ}(t)|$ at its negative minimum if compared to $\sigma=0$. Interestingly, with the range $[0,0.5]$ considered, $\sigma=0.5$ corresponds to the slowest switching [see Fig.~\ref{fig:4}(d)]. This fact may explain why the peculiarity of $\sigma_c=0.5$ occurs at least for $0 \leq \sigma \leq 0.5$. We also find that from Fig.~\ref{fig:4}(d) wherein $\sigma=0.5$ the time period associated with the initial-stage exponential decay is relatively short. This suggests strong scattering between phonons and DBs, which hinders the formation of an exponential decay during this first stage. On the other hand, the second-stage exponential decay seems still available, which may indicate that DBs would likely serve as predominant energy carriers after the phonon-DBs scattering as the foundation of exponential decay of $\rho_Q(0,t)$ should be attributed to DBs.
\begin{figure}[!t]
\vskip-0.2cm
\includegraphics[width=8.8cm]{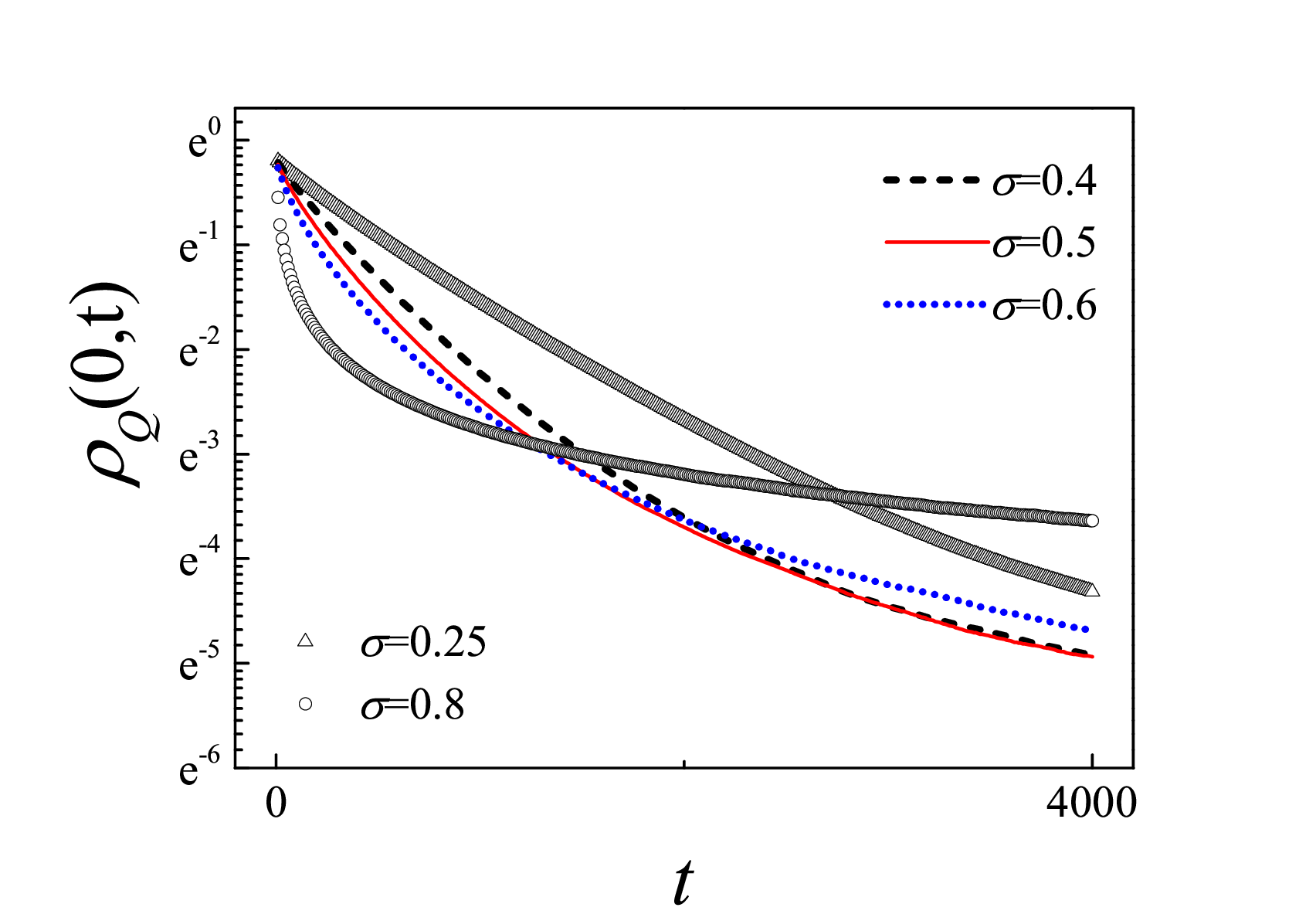}
\vskip-0.3cm
\caption{$\rho_Q(0,t)$ versus $t$ in a single-log plot showing the detailed variation around $\sigma=0.5$.} \label{fig:5}
\end{figure}

We now turn to the results on the vicinity of $\sigma_c=0.5$, as shown by a single-log plot of $\rho_Q(0,t)$ versus $t$ in Fig.~\ref{fig:5}. We note again that a log-log plot would be more helpful to examine the long-time behavior, while from Fig.~\ref{fig:1} the antipersistent energy current correlations only happens at short times, we therefore choose to focus on a single-log plot here. From Fig.~\ref{fig:5} one may identify the special difference of $\sigma=0.5$, if compared to the results obtained for $\sigma=0.4$ and $\sigma=0.6$. Firstly, based on the current data, while all values of $\rho_Q(0,t)$ for $\sigma=0.4$ for different times are always larger than those for $\sigma=0.5$; for $\sigma=0.6$, initially we see a smaller $\rho_Q(0,t)$, but in a long time it will be reversed and such reversion becomes even more prominently for $\sigma=0.8$. Secondly, beyond $\sigma=0.5$, the inital-stage exponential decay exhibited by $\rho_Q(0,t)$ seems almost disrupted and no longer observable (see the result of $\sigma=0.8$). In particular, there exhibits a long tail for $\sigma=0.8$, instead. Based on this long tail, we further conjecture that unlike $0 \leq \sigma \leq 0.5$, for $\sigma>0.5$ phonons may dominate over DBs after undergoing phonon-DBs scattering due to their stronger strength. If this is the case, the relaxation process induced by phonons (perturbed by DBs) would follow a distinct behavior if compared to solely considering DBs. Therefore, even though an antipersistent $C_{JJ}(t)$ can still be observed in certain cases within $\sigma >0.5$ (as evidenced from Fig.~\ref{fig:1}), as $\sigma$ increases further, the long tail of $\rho_Q(0,t)$ for $\sigma>0.5$ would become more evident (see the result of $\sigma=0.8$). Eventually, $\sigma_c=0.5$ may serve as a change point characterizing whether DBs or phonons act as primary surviving energy carriers after phonon-DBs scattering. This may give a full story for why $\sigma_c=0.5$ exhibits peculiar behavior in the whole range of $\sigma \in [0, 1]$, and our findings depicted in Figs.~\ref{fig:4} and~\ref{fig:5} seem to support this conjecture.
\begin{figure}[!t]
\vskip-0.2cm
\includegraphics[width=8.8cm]{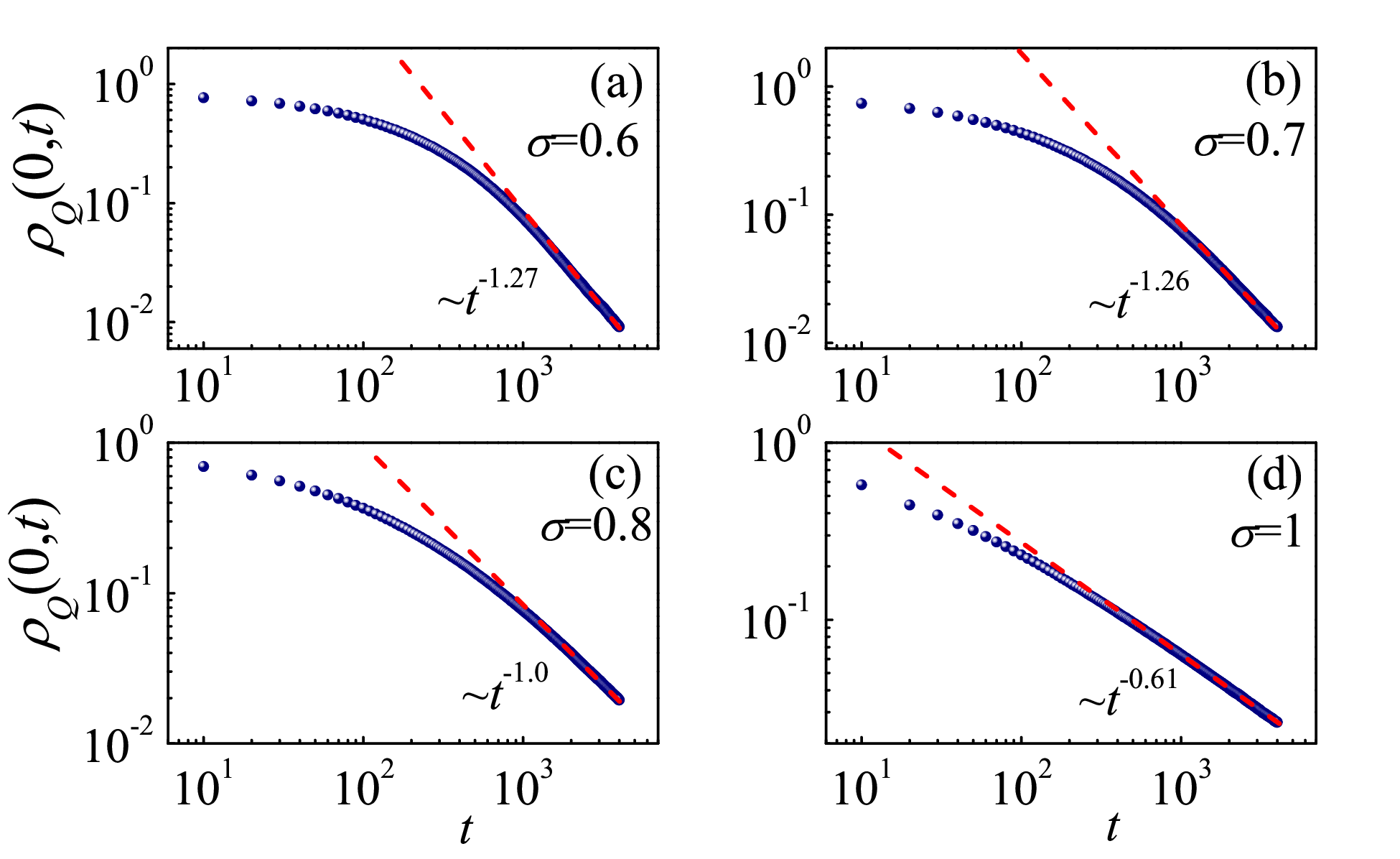}
\vskip-0.3cm
\caption{$\rho_Q(0,t)$ versus $t$ in a log-log plot showing the detailed power-law decay in long times for (a) $\sigma=0.6$, (b) $\sigma=0.7$, (c) $\sigma=0.8$, and (d) $\sigma=1$. } \label{fig:6}
\end{figure}

To capture the observed long tail exhibited in $\rho_Q(0,t)$ for $\sigma>0.5$, similar to $\sigma=0.8$, Fig.~\ref{fig:6} presents a log-log plot of $\rho_Q(0,t)$ versus $t$, providing detailed information for several typical $\sigma$ within $\sigma>0.5$. Since at present only a long time up to $t=4000$ is available, we should admit that to explore the long-time asymptotic behaviors for a $\sigma$ value close to $\sigma_c=0.5$ is still challenging, but the results of Fig.~\ref{fig:6} do provide further insights. As can be seen in Fig.~\ref{fig:6}, based on the current data, while the best fittings of $1/\gamma$ for $\sigma=0.6$ and $\sigma=0.7$ exceed 1 [see Figs.~\ref{fig:6}(a) and (b)], only for $\sigma \geq 0.8$ [see Fig.~\ref{fig:6}(c)] does $1/\gamma$ approximately equal to $1/\gamma \simeq 1$, which now falls within the theoretical prediction range of $1/2 \leq 1/\gamma \leq 1$ according to the L\'{e}vy walk model~\cite{Report2015}. Furthermore, as $\sigma$ increases further, one obtains an exponent close to the theoretical prediction of $\sigma=5/3$ for short-range FPUT systems with NN coupling only~\cite{NNHeatreview-1,NNHeatreview-2,NNHeatreview-3}, with $1/\gamma \simeq 0.61$. Such findings are consistent with the transition from localization to delocalization as shown in Fig.~\ref{fig:2}. Additionally, they indicate that apart from considering $\gamma_c=0.5$, there is a $\gamma$ ($\gamma>0.5$) value, probably being $\sigma \simeq 0.8$ (within the time range considered), above which, the long tail will emerge and the antipersistent energy current correlations will be absent (see also Fig.~\ref{fig:1}).
\section{Absence of antipersistent energy current correlation}
\begin{figure}[!t]
\vskip-0.2cm
\includegraphics[width=8.8cm]{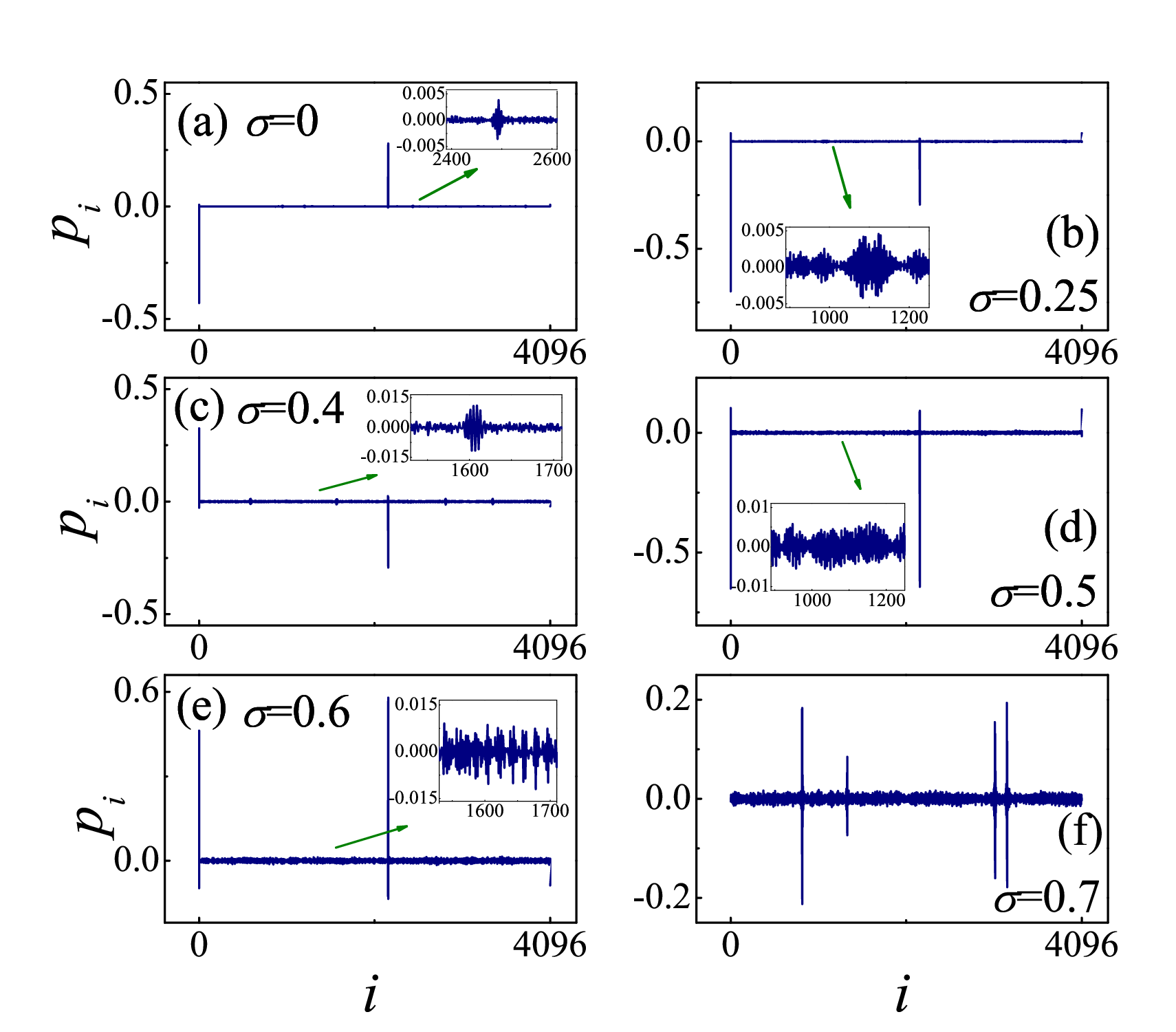}
\vskip-0.3cm
\caption{Snapshots of particle's momentum $p_i$ versus $i$ at a given long time $t=7000$, after initially applying two kicks at $i=1$ and $i=2205$ with momenta $p_1=-0.7$ and $p_{2205}=0.7$ (corresponding to a averaged temperature of $T=0.5$), respectively. The insets in (a)-(e) are a zoom for the typical moving excitations for the given $\sigma$. } \label{fig:7}
\end{figure}
We finally try to give a qualitative explanation for why the absence of antipersistent energy current correlation occurs. In order to achieve this, we examine the phonon-DBs scattering in a zero-temperature system within the whole range of $0\leq \sigma \leq1$. This scattering dynamics have been employed for demonstrating the absence of subdiffusive transport and antipersistent energy current correlation in the mean-field system with truncated long-range interactions, wherein dominant moving excitations emerge when the truncation is sufficiently strong~\cite{LRHeat-ours-2}. Here, based on our analysis presented in Secs.~III and IV, it has become evident that the emergence of the long tail happens at a value of $\sigma>0.5$, which appears to be closely linked to the absence of antipersistent energy current correlation. Therefore, conducting further dynamic examinations may provide valuable insights and enhance our understanding regarding its origin.

To capture the phonon-DBs scattering dynamics in our systems, we perform such a numerical experiment: at time $t=0$ when all particles are at their equilibrium positions (the system is not yet thermalized), we first apply two kicks at the locations of $i=1$ and $i=2205$ with momenta $p_1=-0.7$ and $p_{2205}=0.7$, respectively (it may correspond to an averaged temperature close to $T=0.5$). We then carefully observe the evolution of the dynamics of the system. A visually engaging animation illustrating the phonon-DBs scattering dynamics for various $\sigma$ values with a time lag of $\delta t=10$, displaying momentum $p_i$ versus $i$ for all particles, is given in the Supplementary Material (SM)~\cite{SM}. The relevant snapshots at a long time $t=7000$ are depicted in Fig.~\ref{fig:4}. Before going to discussion of the details, we should note that such a numerical experiment is based on certain specific initial conditions of our setup. It cannot ensure that some of the nonlinear excitations like DBs are sensitive to other initial conditions. In fact, it has been suggested that the stability and mobility of DBs in similar numerical experiments would crucially depend on initial conditions~\cite{DBsMobile-1, DBsMobile-2}. Therefore, our following results can only provide a qualitative but still interesting explanation. Indeed, as evident from both SM and Fig.~\ref{fig:4}, these scattering dynamics suggest that the excitations in these systems can be mainly attributed to phonons and two types of DBs (standing and moving), while for a value of $\sigma>0.5$, probably being $\sigma \simeq 0.7$ [see Fig.~\ref{fig:7}(f)], we identify that the standing DBs can be converted into moving ones. This conversion naturally and qualitatively explains the origin of the long tail and the absence of antipersistent energy current correlations. It is also consistent to the observations of Fig.~\ref{fig:1} and Fig.~\ref{fig:6}(c), wherein a similar value of $\sigma>0.5$ being $\sigma \simeq 0.8$ has been revealed. As to the slight difference between the values of $\sigma \simeq 0.8$ and $\sigma \simeq 0.7$, it is interesting to note that previously in Fig.~\ref{fig:1} and Fig.~\ref{fig:6}(c), we considered a finite-temperature system, while here we deal with a zero-temperature system. In this respect, applying kicks with momentum $|0.7|$ in the latter may not capture all information about a thermalized system in the former, even though one might imagine that for a thermalized system with $T=0.5$, all particles are subjected to many kicks with an average momentum magnitude $|0.7|$ but exhibiting fluctuations since $\sqrt{0.5} \approx 0.7$. This demonstrates again that the revealed absence of antipersistent energy current correlations and the microscopic phonon-DBs scattering dynamics is only qualitatively correlated. In addition to this central information, we are also interesting to find that the movie in SM~\cite{SM} reveals an additional distinct characteristic of $\sigma_c=0.5$ when compared to other values of $\sigma$, i.e., the presence of standing DBs with the lowest frequency. This additional property may further emphasize the peculiarity associated with $\sigma_c=0.5$. However, we acknowledge that currently, we are unable to gain more evidence based solely on the scattering dynamics if compared to the more comprehensive observations provided in Sec. IV.

\section{Conclusion}
To summarize, we have studied the thermal transport properties of the Fermi-Pasta-Ulam-Tsingou (FPUT) model under long-range (LR) interactions, particularly within the challenging strong LR regime (where $0\leq \sigma \leq 1$) that eludes current analytical methods. Our findings uncover the intriguing properties of thermal transport in this regime. Specifically, we observed that the energy current correlation is predominantly antipersistent and varies significantly with $\sigma$, exhibiting nonmonotonic behavior. An important change point emerges at $\sigma_c=0.5$, where this correlation reaches its minimum negative value. Correspondingly, the heat spread is also dependent on $\sigma$, displaying certain special scalings, separated by $\sigma_c=0.5$. For $\sigma \rightarrow 0$, a two-stage exponential decay of the central peak of heat spread can be observed, while for $\sigma \rightarrow 1$, a power-law decay with a long tail over extended time periods emerges. Interestingly, this long tail seems starting to emerge at a value of $0.5<\sigma \leq 1$, supporting the absence of antipersistent energy current correlations as observed. Intriguingly, all these phenomena appear to be qualitatively linked to the unique dynamics of $\sigma$-dependent phonon-DBs scatterings dynamics. Combining all of these observations, we conjecture $\sigma_c=0.5$ to be a change point characterizing whether final surviving energy carriers after the phonon-DBs scattering are DBs or phonons. In this respect, the absence of antipersistent energy current correlations can also be qualitatively related to the observation of standing DBs converting into moving ones at a value of $0.5<\sigma \leq 1$. All these findings provide a fresh perspective on thermal transport in LR interacting systems that extends beyond mean-field and weak LR regimes, even though further detailed examination of the conjecture is still required.

Regarding the newly identified critical point $\sigma_c=1/2$, its discovery is unexpected, especially given previous knowledge only of the special points at $\sigma=1$~\cite{LRHeat-2016}---the demarcation between strong and weak LR regimes---and $\sigma=2$~\cite{LRHeat-2017-1,LRHeat-ours-1}, where quasi-integrable or weakly nonintegrable dynamics occur. This finding could be deeply connected to similar behaviors noted in chaos suppression~\cite{add-1,add-2} and universal dynamics thresholds~\cite{add-3} in relevant LR interacting systems, presenting an intriguing avenue for future study.

\begin{acknowledgments}
We are grateful to the referees for their valuable comments, which help us improve the presentation. D.X. is supported by NNSF (Grant No.~12275116) of China, NSF (Grant No.~2021J02051) of Fujian Province of China, and the start-up fund of Minjiang University; J.W. is supported by NNSF (Grant No.~12105122) of China.
\end{acknowledgments}


\end{document}